\documentclass[onecolumn,showpacs,preprintnumbers,amsmath,amssymb]{revtex4-2}

\usepackage{graphicx, amsmath, amssymb, amsfonts, mathtools, mathrsfs, color}
\usepackage{natbib, comment} 
\usepackage[justification=RaggedRight]{caption}
\usepackage[margin = 1.5cm]{geometry} 
\usepackage{xcolor}
\usepackage[capitalise]{cleveref}

\newcommand{\pd}[2]{ \frac{ \partial #1}{ \partial #2 } }

\newcommand{\bvec}[1]{{\mathbf{#1}}}
\newcommand{\grad}{\nabla} 
\newcommand{\abs}[1]{\left| #1 \right|}

\newcommand{\uu}{\bvec{u}}
\newcommand{\uhat}{\hat{\uu}}
\newcommand{\vv}{\bvec{v}}
\newcommand{\Bvec}{\bvec{B}}
\newcommand{\Bhat}{\hat{\Bvec}}
\newcommand{\Avec}{\bvec{A}}
\newcommand{\Ahat}{\hat{\Avec}}
\newcommand{\jvec}{\bvec{j}}
\newcommand{\xx}{\bvec{x}}
\newcommand{\kk}{\bvec{k}}
\newcommand{\nn}{\bvec{n}}

\newcommand{\kset}{\mathcal{S}[{\ksq}]}
\newcommand{\colvec}[3]{\begin{bmatrix} #1 \\ #2 \\ #3 \end{bmatrix}}

\newcommand{\Rm}{\text{Rm}}
\newcommand{\vd}[2]{\frac{\delta #1 }{\delta #2}}
\newcommand{\Vol}{\text{Vol}(\Omega)}
\newcommand{\uopt}{\uu_{\text{opt}}}
\newcommand{\uprov}{\uu_{\text{prov}}}
\newcommand{\vrand}{\uu_{\text{rand}}}
\newcommand{\weight}{w}
\newcommand{\Const}{{C}}
\newcommand{\proj}{ \mathcal{P}_{\text{df}} }
\newcommand{\phat}{\hat{\mathcal{P}}_{\text{df}}}
\newcommand{\ksq}{k^2_{\text{max}}}
\newcommand{\nmodes}{N_{\text{modes}}}
\newcommand{\Mdot}{\dot{M}}

\newcommand{\MdotOpt}{\dot{M}_{\text{opt}}}
\newcommand{\Bopt}{\Bvec_{\text{opt}}}
\newcommand{\jopt}{\jvec_{\text{opt}}}

\newcommand{\ex}{\bvec{e}_1}

\newcommand{\ez}{\bvec{e}_3}

\def\Xint#1{\mathchoice
   {\XXint\displaystyle\textstyle{#1}}%
   {\XXint\textstyle\scriptstyle{#1}}%
   {\XXint\scriptstyle\scriptscriptstyle{#1}}%
   {\XXint\scriptscriptstyle\scriptscriptstyle{#1}}%
   \!\int}
\def\XXint#1#2#3{{\setbox0=\hbox{$#1{#2#3}{\int}$}
     \vcenter{\hbox{$#2#3$}}\kern-.5\wd0}}

\def\dashint{\Xint-}

\begin{document}
\title{Optimal velocity fields for instantaneous magnetic-field growth}

\author{Nicholas J.~Moore }
\email{nickmoore83@gmail.com}
\affiliation{Department of Mathematics, Colgate University, Hamilton, NY 13346, USA}
\author{Stefan G. Llewellyn Smith}
\email{sgls@ucsd.edu}
\affiliation{Department of Mechanical and Aerospace Engineering, Jacobs School of Engineering, UCSD, La Jolla CA 92093-0411, USA, Scripps Institution of Oceanography, UCSD, La Jolla, CA 92093-0213, USA}

\begin{abstract}
We consider a variant of the kinematic dynamo problem. Rather than prescribing a velocity field and searching for high-growth magnetic fields via an eigenvalue problem, we treat the seed magnetic-field structure as given and ask which velocity field maximally enhances its instantaneous growth. We show this second problem has an elegant formulation in terms of variational calculus. Upon simultaneously constraining the velocity's kinetic energy and enstrophy, the Euler-Lagrange equation leads to a forced Helmholtz partial differential equation for the optimal velocity field. For the special case of fixed kinetic energy and unconstrained enstrophy, the optimal velocity field everywhere opposes the divergence-free projection of the Lorentz force. In the more general setting, the optimal velocity field can be found through numerical solution of the forced Helmholtz equation. We construct 2.5-dimensional numerical examples to support the theoretical findings, and then leverage the newly found optimal velocity fields to accelerate numerical optimization of the magnetic-velocity field pair for maximal instantaneous growth rate.
\end{abstract}
\maketitle

\section{Introduction}

The flow of conducting fluids in the interiors of stars and planets creates large-scale magnetic fields through the dynamo mechanism  \cite{babcock1961topology, leighton1969magneto, roberts1972dynamo, parker1993solar, brandenburg2005astrophysical, Moffatt2019, Tobias2021}. Most often, the interior flows are driven by thermal convection in conjunction with other effects such as rotation, shear-driven instabilities, and oblateness \cite{gilman1969rossby, balbus1994stability, spruit2002dynamo, camassa2012stratified, roberts2013genesis, matilsky2023stellar, matilsky2024solar, petitdemange2024tayler, vasil2024solar}. Furthermore, reversals of the magnetic dipole, such as those that occur irregularly in the earth \cite{bullard1950westward, Glatzmaier1998, constable2006earth, petrelis2009simple, Yadav2016, jones2025low} and regularly in the sun \cite{charbonneau2014solar, hathaway2015solar, matilsky2020exploring, matilsky2024solar}, may be tied to reorganization of the flow's large-scale circulation \cite{moore2024large, zhang2025low}, highlighting the inextricable link between magnetic fields and the underlying velocity fields.

A natural question is: What types of flow fields can produce dynamo action and which flows are best at it? Kinematic dynamo theory formulates this question by prescribing a large-scale velocity field, often steady in time, and determining the growth rate of a corresponding magnetic field through the solution of an eigenvalue problem \cite{Moffatt2019, Tobias2021}.
The eigenfunction with the largest eigenvalue corresponds to the magnetic field with the highest growth rate for the particular velocity field that was prescribed. A class of recent studies have extended the kinematic dynamo problem beyond the simple eigenvalue formulation. They apply variational techniques to numerically optimize the velocity-magnetic-field pair for long-term magnetic-field growth \cite{Willis2012, Chen2015, Chen2018, herreman2018minimal, Luo2020}. These studies have considered a variety of domain geometries, including triply-periodic \cite{Willis2012}, cubic \cite{Chen2015}, and spherical \cite{Chen2018, Luo2020} domains.

Here, we propose a new variant of the kinematic dynamo problem. Rather than prescribing the velocity field, we treat the seed magnetic field as the prescribed object. We then seek the companion velocity field that maximally enhances its instantaneous growth rate. We find that this problem has an elegant formulation in terms of variational calculus. Simultaneously constraining the flow's kinetic energy and enstrophy results in a forced Helmholtz partial differential equation (PDE) for the optimal velocity field. In this PDE, the forcing term is proportional to the divergence-free projection of the Lorentz force. For the special case of fixed kinetic energy and no constraint on enstrophy, the problem is readily solvable and the optimal velocity field everywhere opposes the divergence-free projection of the Lorentz force. That is, the optimal velocity field maximizes the work done against the Lorentz force \cite{Moffatt2019}, subject to the constraints placed on the flow. In the more general case of jointly constrained kinetic energy and enstrophy, the optimal velocity field can be found through numerical solution of the forced Helmholtz equation.

The current paper explores these new ideas. Section \ref{sec:opt_fields} establishes the governing equation and formulates the optimization problem in terms of variational calculus. The variational problem gives rise to the forced Helmholtz equation whose solution yields the optimal velocity field $\uopt$ corresponding to the particular magnetic field that was prescribed. In Section \ref{sec:numerical_examples} we construct numerical examples to demonstrate optimality. For simplicity, our numerical examples involve 2.5-dimensional magnetic and velocity fields in  doubly-periodic domains. These fields possess sufficient complexity to produce strong {\em instantaneous} magnetic-field growth, which is the main interest of this paper, even though long-time dynamos are ruled out by anti-dynamo theorems \cite{Moffatt2019, Tobias2021}. Through comparison against randomly selected velocity fields, we demonstrate that $\uopt$ always produces the optimal magnetic-energy growth rate for the seed magnetic field that was prescribed. Leveraging the efficient method for determining $\uopt$, we then perform numerical optimization over the space of possible magnetic fields. This process yields the optimal magnetic-velocity field pair $(\Bvec,\uu)$ in the assigned function space. We close with a discussion in Section \ref{sec:discussion}.

\section{Optimal velocity fields}
\label{sec:opt_fields}

\subsection{Governing equations}
Given a velocity field $\uu$, the solenoidal magnetic field $\Bvec$ evolves according to the induction equation \cite{Brandenburg1998, Moffatt2019, Tobias2021}, given in dimensionless form by
\begin{align}
\label{induction}
& \pd{}{t}  \Bvec = \grad \times \left( \uu \times \Bvec \right) + \Rm^{-1} \grad^2 \Bvec  \, , \\
\label{solenoid}
& \grad \cdot \Bvec = 0 \, .
\end{align}
Above, the magnetic Reynolds number is $\Rm = UL/\eta$, where $\eta$ is the magnetic diffusivity of the fluid, $U$ is a characteristic velocity, and $L$ is a characteristic length scale, assumed to be the same for $\Bvec$ and $\uu$.
In our nondimensionalization, we choose units in which the magnetic permeability $\mu_0$ is unity, so that the current is simply the curl of the magnetic field, $\jvec = \grad \times \Bvec$ \cite{brandenburg2005astrophysical}.

The dimensionless, volume-averaged magnetic energy over domain $\Omega$ is given by
\begin{equation}
\label{Mdefn}
M(t) = \frac{1}{2} \, \dashint_{\Omega} \abs{\Bvec}^2 \, dV \, ,
\end{equation}
where the dashed integral indicates the mean value over  $\Omega$,
\begin{equation}
\dashint_{\Omega} \cdot  \,\, dV := \frac{1}{\Vol} \int_{\Omega} \cdot \,\, dV \, .
\end{equation}
The rate of change of magnetic energy can be calculated as
\begin{align}
\label{Mdot1}
\Mdot = \dashint_{\Omega} \Bvec \cdot \Bvec_t \, dV 
 = \dashint_{\Omega} \Bvec \cdot \left( \grad \times \left( \uu \times \Bvec \right) + \Rm^{-1} \grad^2 \Bvec \right) \, dV \, .
\end{align}
Integrating by parts and neglecting surface terms yields the alternate form \cite{Tobias2021},
\begin{align}
\label{Mdot2}
\Mdot = - \, \dashint_{\Omega} \uu \cdot \left( \jvec \times \Bvec \right) \, dV - \Rm^{-1} \dashint_{\Omega} \abs{\jvec}^2 \, dV \, .
\end{align}
Surface terms may alter \cref{Mdot2} based on the boundary conditions placed on the magnetic field. Unlike velocity-field boundary conditions, which tend to be straightforward and uncontroversial, magnetic-field boundary conditions typically involve some degree of simplification \cite{roberts1988future, charbonneau2014solar, matilsky2020exploring, Tobias2021}. However, insight into the essential physics can be gained by considering only the volume terms shown in \cref{Mdot2} and neglecting any surface contributions. For this reason, the remainder of the paper focuses on maximizing $\Mdot$ given by \cref{Mdot2}.

\subsection{Optimization problem}

We now seek to maximize the growth rate $\Mdot$ of the magnetic field. The key insight is that while \cref{Mdot2} is nonlinear in the magnetic field through the presence of $\Bvec$ and $\jvec$, it is linear in the velocity $\uu$. Hence, prescribing the spatial structure of $\Bvec$ and searching for the optimal $\uu$ is a tractable problem which we will show can be solved by variational calculus, reminiscent of several recently examined problems in fluid mechanics \cite{hassanzadeh2014wall, tobasco2017optimal, Rajkotia-Zaheer_Goluskin_2026}.
The field $\uu$ is subject to the constraints of a physical velocity field, including: (1) incompressibility $\grad \cdot \uu = 0$; (2) bounded kinetic energy $E$; and (3) bounded enstrophy $\mathcal{E}$. To satisfy incompressibility, we enforce the equivalent condition:
\begin{align}
\label{div_free_var}
\dashint_{\Omega} \Pi(\xx) \, \left(\grad \cdot \uu \right) \, dV = 0 \, ,
\end{align}
for any sufficiently smooth $\Pi(\xx)$. Later, $\Pi(\xx)$ will become a Lagrange multiplier in the optimization problem.

Meanwhile, the dimensionless, volume-averaged kinetic energy and enstrophy of the velocity field $\uu$ are given by
\begin{align}
\label{kinetic_energy}
E &= \frac{1}{2} \, \dashint_{\Omega} \abs{\uu}^2 \, dV \, , \\
\label{enstrophy}
\mathcal{E} &= \frac{1}{2} \, \dashint_{\Omega} \abs{\grad \times \uu}^2 dV \, .
\end{align}

In optimizing the velocity field, it is necessary to constrain the search to a normalized space; otherwise, the optimization  will simply seek velocity fields of ever greater energy and enstrophy. Here, we normalize velocity fields by a weighted average of energy and enstrophy. That is, we enforce
\begin{align}\
\label{energy_enstrophy}
& \weight E + (1-\weight) \mathcal{E} = 1 \, ,
\end{align}
where $w$ weights the relative contributions of energy and enstrophy. As long as $0 < w < 1$, \cref{energy_enstrophy} ensures that both energy and enstrophy are bounded. Hence, it typically suffices to set $w = 1/2$. The edge cases of $w=0$ and $w=1$ merit brief attention: $w=1$ corresponds to unit energy and no constraint on enstrophy, while $w=0$ corresponds to unit enstrophy and no constraint on energy. As noted by Willis (2012) \cite{Willis2012}, bounding only the energy while leaving the enstrophy unconstrained can result in optimal velocity fields with unphysically sharp gradients. It is for this reason that we most often bound both energy and enstrophy by taking $0<w<1$.

The velocity field $\uu$ that maximizes \cref{Mdot2} subject to constraints \cref{div_free_var} and \cref{energy_enstrophy} must satisfy the Euler-Lagrange equations. After applying integration by parts to \cref{div_free_var} and taking the variational derivative with respect to $\uu$, the Euler-Lagrange equations become
\begin{equation}
\label{EulerLagrange}
\vd{\Mdot}{\uu} - \grad \Pi + \lambda \weight \vd{E}{\uu} + \lambda (1-\weight) \vd{\mathcal{E}}{\uu}  = 0 \, ,
\end{equation}
for some $\Pi(\xx)$ and $\lambda$.
The function $\Pi(\xx)$ and the scalar $\lambda$ are Lagrangian multipliers that will be selected to enforce the incompressibility constraint, $\grad \cdot \uu = 0$, and the energy-enstrophy constraint, \cref{energy_enstrophy}, respectively.

\subsection{The optimal velocity field}

In \cref{EulerLagrange}, it is straightforward to calculate the variational derivative $\delta \Mdot/\delta \uu$ using \cref{Mdot2} and $\delta E/\delta \uu$ using \cref{kinetic_energy}. These derivatives are given by
\begin{align}
\vd{\Mdot}{\uu} = - \jvec \times \Bvec \, , \quad
\vd{E}{\uu} = \uu \, .
\end{align}
The variational derivative of enstrophy $\mathcal{E}$ requires only slightly more finesse. Taking the variation of \cref{enstrophy} and integrating by parts gives
\begin{align}
\label{dEnstrophy1}
\delta \mathcal{E} = 
& \dashint_{\Omega} (\grad \times \uu) \cdot (\grad \times \delta \uu) \, dV = \\
\label{dEnstrophy2}
&- \dashint_{\Omega} \grad^2 \uu \cdot \delta \uu \, dV - 
\dashint_{\partial \Omega} \left( \left( \grad \times \uu \right) \times \delta \uu \right) \cdot \nn \, dS \, .
\end{align}
As before, we will neglect the surface contribution in \cref{dEnstrophy2}. We note this assumption is justified if $\uu$ satisfies either no-slip or stress-free boundary conditions.
After dropping the surface term, the variational derivative of enstrophy becomes
\begin{align}
\vd{\mathcal{E}}{\uu} = - \grad^2 \uu \, .
\end{align}

With all of the variational derivatives known, the Euler-Lagrange equation (\ref{EulerLagrange}) takes the form
\begin{align}
\label{PDE1}
\weight \uu - (1-\weight) \grad^2 \uu = \Const \, \left( - \jvec \times \Bvec - \grad \Pi \right) \, ,
\end{align}
where we have set $\Const = -1/\lambda$. Recall that $\Pi(\xx)$ is selected to enforce the incompressibility condition, $\grad \cdot \uu = 0$. Since the divergence commutes with the Laplacian, we can equivalently choose $\Pi(\xx)$ to render the right-hand-side of \cref{PDE1} divergence free, $\grad^2 \Pi = \grad \cdot (- \jvec \times \Bvec)$. To this end, we introduce the {\em divergence-free projection operator} $\proj$ that acts on any sufficiently smooth vector field $\vv$ by
\begin{align}
\label{div_free_project}
\proj[ \vv ] = \vv - \grad p \, ,  \text{  where } \grad^2 p = \grad \cdot \vv \, .
\end{align}
It follows that $\grad \cdot \proj[ \vv ] = 0$ for any such field $\vv$.

Applying $\proj$ to \cref{PDE1} yields the following PDE for $\uu = \uopt$:
\begin{align}
\label{PDE2}
\weight \uu - (1-\weight) \grad^2 \uu = \Const \, \proj \left[ - \jvec \times \Bvec \right]
&&\mbox{\em forced Helmholtz PDE} \, ,
\end{align}
where $\Const$ is chosen to enforce the energy-enstrophy constraint \cref{energy_enstrophy}. The above is a {\em forced Helmholtz PDE} for the velocity field $\uu = \uopt$ that maximizes the instantaneous growth of the given $\Bvec$. The Lorentz force is given by $\jvec \times \Bvec$, and therefore the forcing term in \cref{PDE2} is proportional to the {\em divergence-free projection of the negative Lorentz force}.
Furthermore, thanks to the positive-semidefiniteness of $-\grad^2$, the differential operator on the left-hand-side of \cref{PDE2} is non-singular as long as $0<w<1$; i.e.~it is a non-singular Helmholtz operator. Owing to these properties, one may choose any sufficiently smooth solenoidal magnetic field $\Bvec$ and the solution of \cref{PDE2} yields the velocity field that maximizes its instantaneous growth under evolution by \cref{induction}.

A few special cases merit brief consideration. First, setting $\weight=1$ corresponds to fixing $E=1$ but leaving $\mathcal{E}$ completely unconstrained.  
In this case \cref{PDE2} is simply an algebraic equation for the optimal velocity field, 
\begin{align}
\uopt = \Const \, \proj \left[ - \jvec \times \Bvec \right]
&&\mbox{\em fixed energy and unconstrained enstrophy} \, ,
\end{align}
that is, the divergence-free projection of the negative Lorentz force. This velocity field maximizes the growth rate of $\Bvec$ by maximizing the work done against the Lorentz force \cite{Moffatt2019}, subject to the constraints of incompressibility and fixed energy. This field may, however, possess arbitrarily large enstrophy. Taking the opposite extreme of $w=0$ fixes the enstrophy while leaving energy unconstrained. In this case, \cref{PDE2} takes the form of a Poisson equation
\begin{align}
\grad^2 \uu = \Const \, \proj \left[ \jvec \times \Bvec \right]
&&\mbox{\em fixed enstrophy and unconstrained energy} \, .
\end{align}
Depending on boundary conditions, however, the Poisson operator may potentially be singular. For these reasons, we typically choose $0<w<1$.



Returning to the general case of \cref{PDE2}, we comment on the sign of the constant $\Const$ used to enforce the energy-enstrophy constraint, \cref{energy_enstrophy}. Taking $\Const$ to be the unique {\em positive} constant satisfying \cref{energy_enstrophy} yields the maximum possible value of $\Mdot$ in \cref{Mdot2}, while taking the equivalent negative constant yields the minimum $\Mdot$. Due to the linearity of \cref{Mdot2} in $\uu$, the minimum value of $\Mdot$ is simply the negative of the maximum value. Therefore, in the absence of magnetic diffusion ($\Rm^{-1}=0$), the solution $\uu=\uopt$ of \cref{PDE2} with $\Const >0$ yields a non-negative growth rate, $\Mdot \ge 0$. In the presence
of magnetic diffusion ($\Rm^{-1} > 0$), the maximal $\Mdot$ may be negative depending on the relative strength of dissipation, but the solution to \cref{PDE2} is still guaranteed to be the maximizer since the dissipation term in \cref{Mdot2} is {\em independent} of $\uu$.

\section{Numerical Examples}
\label{sec:numerical_examples}

We now construct numerical examples to illustrate essential features of the optimizing velocity field $\uopt$. For simplicity, we consider only periodic domains here, with plans to extend the experiments to other geometries, such spherical ones, in the future.
The numerical examples serve two main purposes:
\begin{enumerate}
\item For an arbitrarily prescribed magnetic field $\Bvec$, we will verify that the numerical solution of \cref{PDE2} does indeed maximize the instantaneous growth rate $\Mdot$ by comparing against growth rates produced by the same $\Bvec$ but randomly sampled velocity fields.
\item We will leverage the numerical solution of \cref{PDE2} to facilitate numerical optimization of the prescribed $\Bvec$ for overall maximal $\Mdot$. Compared to direct optimization of the pair $(\Bvec, \uu)$, this new approach reduces the dimension of the optimization space by roughly half.
\end{enumerate}

Our aim is to construct numerical examples that illuminate essential features of magnetic-field growth with minimal complexity. Two-dimensional vector fields --- that is, two vector components depending on two spatial variables --- are known to be too simple to capture essential magneto-hydrodynamic features \cite{Moffatt2019, Tobias2021}. Meanwhile, fully three-dimensional vector fields --- three vector components depending on three spatial variables --- possess many degrees of freedom, even in the case of severely band-limited fields. This complexity places a practical limit on the ability to sample the corresponding function spaces.
In an effort to balance simplicity and complexity, we consider so-called 2.5-dimensional vector fields, in which 3 vector components depend on 2 spatial variables \cite{smith2004vortex}. In our tests, both the magnetic fields and velocity fields are 2.5-dimensional. It is important to note that a magnetic field depending on only two spatial variables cannot be sustained {\em in long time} by dynamo action \cite{jones2008course, Tobias2021}. However, the complexity is sufficient to generate strong {\em instantaneous} growth of the magnetic energy, which is the purpose of the present paper, and simplicity compared to fully three-dimensional fields offers greater interpretability.

In all numerical experiments, we set the magnetic diffusivity to zero, $\Rm^{-1}=0$. Recall that non-vanishing magnetic diffusivity, $\Rm^{-1}>0$, generally reduces $\Mdot$, but does not alter the optimizing field $\uopt$ since the dissipative term in \cref{Mdot2} does not depend on velocity.

\subsection{Numerical algorithms}

\subsubsection{Background}

In our tests, we assume a two-dimensional, doubly-periodic domain $\Omega = [0,2\pi)^2$. For a given non-negative integer $\ksq$, consider a real-valued vector field $\Avec$ composed of Fourier modes with square wavenumber up to $\ksq$, i.e. $\abs{\kk}^2 = k_1^2+k_2^2 \le \ksq$. Such a vector field can be represented through its Fourier series as
\begin{align}
\label{FourierSeries}
\Avec(\xx) = \left( \frac{1}{2} \Ahat_\bvec{0} + \sum_{\kk \in \kset} \Ahat_\kk \exp \left( i \kk \cdot \bvec{x} \right) \right) \,\,+\,\, C.C.,
\end{align}
where $\xx = (x_1,x_2) \in \Omega$ and $\mathcal{F}(\Avec)_\kk = \Ahat_\kk$ represents the Fourier coefficient corresponding to wavevector $\kk = (k_1,k_2)$. The notation C.C.~indicates that the complex conjugate should be added to render the field $\Avec$ real valued.
The set 
$\kset = \{ (k_1,k_2) \in \mathbb{Z}^2 \,|\, k_1^2+k_2^2 \le \ksq \text{ and } 
(k_1 > 0 \text{ or } (k_1=0 \text{ and } 0 < k_2 \le k_{\text{max}}))
 \}$ is a minimal set of wavevectors, which, upon adding the complex conjugate, is capable of representing an arbitrary real field with maximum square wavenumber $\ksq$.
The Fourier coefficients are given by the usual formula,
\begin{equation}
\label{FourierModes}
\mathcal{F}({\Avec})_\kk =
\Ahat_\kk = 
\dashint_{\Omega} \Avec(\xx) \exp(-i \kk \cdot \xx) \,\, d\xx \, .
\end{equation}
As indicated above, we will use the hat notation for Fourier coefficients, but may also use the operator notation $\mathcal{F}$ for particularly long expressions.

Applying Parseval's identity to \cref{kinetic_energy,enstrophy}, the energy and enstrophy of the field $\Avec$ in \cref{FourierSeries} are
\begin{align}
\label{Energy_compute}
&E = \frac{1}{2} \abs{\Ahat_{\bvec{0}}}^2 + 
\sum_{\kk \in \kset} \abs{\Ahat_\kk}^2\, , \\
\label{Enstrophy_compute}
&\mathcal{E} = \sum_{\kk \in \kset} \abs{\kk \times \Ahat_\kk}^2 \, .
\end{align}
Above, and hereafter it is implied that the wavevector should be extended to a three-dimensional vector where necessary, $\kk = (k_1,k_2,0)$.

\subsubsection{Spectral computation of the optimizing velocity field}

We now discuss our method to  solve \cref{PDE2} numerically for the optimal velocity field $\uopt$ that corresponds to a specified 2.5-dimensional magnetic field $\Bvec:\Omega \to \mathbb{R}^3$. We perform all computations directly in spectral space, with no Fourier transforms to the physical domain. Below, $\nmodes$ denotes the number of independent wavevectors in each test, i.e.~the number of wavevectors in the set $\kset$ along with the zero mode. To provide some baseline intuition, \cref{tab:modes} lists values of $\nmodes$ for the first few values of $\ksq$. 

To solve for $\uopt$, it is first necessary to compute the Lorentz force $\jvec \times \Bvec$ that underlies the inhomogeneous term in \cref{PDE2}. It is straightforward to compute the derivative $\jvec = \grad \times \Bvec$ in spectral space. We then compute $\jvec \times \Bvec$ in spectral space by directly summing the $O(\nmodes^2)$ terms in the product of the two Fourier series. We retain all modes in the cross product, and, as a result, the term $\jvec \times \Bvec$ contains more Fourier modes than $\Bvec$.
We point out that, alternatively, a fast Fourier transform (FFT) could be used to compute this term in physical space with $O(\nmodes \log \nmodes )$ operations, i.e.~a pseudo-spectral method. However, we aim to perform sampling and optimization over vector-valued function spaces whose dimension grows rapidly with $\ksq$ (see \cref{tab:modes}). It will therefore only be practical to consider relatively small values of $\ksq$, for which direct spectral methods outperform pseudo-spectral ones \cite{trefethen2000spectral, boyd2001chebyshev}. To give one illustrative example, setting $\ksq=4$ gives $\nmodes=7$ and the space in which $\Bvec$ resides is 25-dimensional, which is already a non-trivial space to sample numerically. The matter gets substantially worse as $\ksq$ increases.

Returning to the numerical solution of \cref{PDE2}, it is necessary to remove the divergence from the Lorentz force $\jvec \times \Bvec$ through the operator $\proj$ defined in \cref{div_free_project}. We introduce $\phat$ as the spectral counterpart to $\proj$, that is, a divergence-free projection operator that acts on Fourier components via
\begin{equation}
\label{div_free_spec}
\phat[\Avec]_\kk = 
\Ahat_\kk - \frac{ \Ahat_\kk \cdot \kk }{\abs{\kk}^2} \colvec
{k_1}
{k_2}
{0}
\qquad \text{for } \kk \ne \bvec{0} \, .
\end{equation}
The constant mode $\Ahat_{\bvec{0}}$ remains unmodified by $\phat$ since it is already divergence free. It is straightforward to verify the $\phat$ renders each Fourier mode of $\Avec$ divergence free and is consistent with \cref{div_free_project}.

Once $\phat \left[ - \jvec \times \Bvec \right]$ is computed, it is straightforward to solve \cref{PDE2} in spectral space. We first set $\Const = 1$ to obtain a provisional solution $\uu = \uprov$, with Fourier components given by
\begin{equation}
\label{prov_soln}
\uhat_\kk^{(\text{prov})} = \frac
{\phat[-\jvec \times \Bvec]_\kk}
{w + (1-w) \abs{\kk}^2} 
\end{equation}
We remark that the denominator, $w + (1-w)\abs{\kk}^2$, is the {\em symbol} of the differential operator from \cref{PDE2} in spectral space. Once the provisional solution is known, it only remains to chose the constant $\Const$ to satisfy the energy-enstrophy condition \cref{energy_enstrophy} with weight $w$. We define the weighted norm of a vector field $\uu$ as
\begin{equation}
\label{wnorm}
||\uu||^2_w = w E[\uu] + (1-w) \mathcal{E}[\uu]
\end{equation}
where $E$ and $\mathcal{E}$ are computed from \cref{Energy_compute,Enstrophy_compute} respectively. Then the optimal velocity field is simply the provisional solution normalized by $|| \cdot ||_w$,
\begin{equation}
\label{uopt_normalized}
\uopt = \frac{\uprov}{||\uprov||_w}
\end{equation}

We note that, in the doubly-periodic domain $\Omega = [0,2\pi)^2$, integration by parts implies that the Lorentz for $\jvec \times \Bvec$ is mean free. \Cref{prov_soln} implies that $\uopt$ is also mean free, implying that its constant Fourier mode vanishes. We also remind the reader that the optimal velocity field $\uopt$ implied by \cref{prov_soln,uopt_normalized} is unaltered by magnetic diffusivity, $\Rm^{-1} \ne 0$, since the dissipation term in \cref{Mdot2} is independent of velocity.

\subsubsection{Sampling divergence-free vector fields spectrally}
\label{sec:sampling}

In the numerical tests below, we will select random magnetic fields and velocity fields by sampling their Fourier coefficients. Each of these fields must satisfy the divergence-free constraint. For $\kk \ne \bvec{0}$, the Fourier coefficients of an arbitrary divergence-free vector field $\Avec$ can be constructed from two complex numbers $z_1$ and $z_2$ as follows
\begin{align}
\label{df_mode}
\Ahat_\kk = \colvec
{ (\sqrt{a} \, k_2/\abs{\kk}) \, z_1}
{-(\sqrt{a} \, k_1/\abs{\kk}) \, z_1}
{z_2}
\end{align}
It is straightforward to verify that each mode $\Ahat_\kk e^{i \kk \cdot \xx}$ is divergence free. Meanwhile, the real constant mode $\Ahat_{\bvec{0}}$ can be constructed from three independent real numbers.

We will sample the complex numbers $z_1$ and $z_2$ independently from identical distributions (i.i.d.).
The parameter $a>0$ in \cref{df_mode} controls, in a statistical sense, the magnitude of the in-plane $(x_1,x_2)$ component versus the out-of-plane $x_3$ component of $\Avec$. For shear fields in which either $k_1=0$ or $k_2=0$, setting $a=1$ gives statistically equal magnitudes of the two components. However, for other modes, different values of $a$ achieve better balance. For the mode $(k_1,k_2)=(1,1)$, for example, $a=2$ produces a statistically equal magnitude of each component. Nonetheless, the entire space of divergence-free vector fields can be represented by any choice of $a>0$. We set $a=1$ since most of our tests use small values of $\ksq$ for which the shear components are statistically important.

Briefly recall the first goal of our numerical tests: For a specified magnetic field, we aim to verify that the theoretically optimal velocity field, $\uopt$, indeed produces the largest possible growth rate $\Mdot$ of magnetic energy. The field $\uopt$ is found by numerically solving \cref{PDE2} and then the theoretically optimal growth rate $\MdotOpt$ is computed using \cref{Mdot2}. To confirm this growth rate is optimal, we compare against $\Mdot$ values produced by the same magnetic field but randomly selected velocity fields $\vrand$, where the random sampling is performed using \cref{df_mode}.
Preliminary tests indicate that sampling the random numbers $z_1$ and $z_2$ from normal distributions produces sampled values of $\Mdot$ whose magnitudes lie substantially below the theoretical $\MdotOpt$. This observation, while it confirms our theoretical result, suggests that the space of velocity fields was not adequately searched. To better search these large spaces, we make a few key modifications to the sampling procedure. First, we sample the magnitudes of the complex numbers $z_1$ and $z_2$ from a heavy-tailed distribution, specifically a student's t-distribution with $\nu=3$. This choice increases the likelihood of extreme events and thus raises the largest sampled magnitude of $\Mdot$. Second, we apply some basic rejection sampling to further increase the sampled values of $\Mdot$. In particular, we accept velocity-field samples with probability proportional to $\exp(2(x^2-1))$ where $x=\Mdot/\MdotOpt$. These two choices increase the representation of velocity fields producing large $\Mdot$, thus offering a more compelling confirmation of the theoretical results.


\subsubsection{Spectral truncation of the optimizing velocity field}
\label{sec:truncation}

Owing to the fact that the cross product $\jvec \times \Bvec$ is quadratic in $\Bvec$, the optimal velocity field $\uopt$ determined by \cref{PDE2} generally contains more Fourier modes than does $\Bvec$. The field $\uopt$ therefore resides in a function space of higher dimension, and if the size of this space is too large it can hinder the practical ability to numerically sample the space.
As such, it may be desirable to determine the optimal velocity field residing in a lower-dimensional subspace, for example the same Fourier space as $\Bvec$. We show here that direct truncation of $\uopt$ to a smaller Fourier space solves the same optimization problem as outlined in \ref{sec:opt_fields} with the additional constraint that the field lies in the specified space.

To begin, consider the optimization problem of Section \ref{sec:opt_fields} with the additional constraints
\begin{equation}
\label{truncate1}
\mathcal{F} (\uopt)_\kk = \bvec{0} \quad \text{for } |\kk|^2 > \ksq \, .
\end{equation}
Through \cref{FourierModes}, these constraints are equivalent to
\begin{equation}
\label{truncate2}
\dashint_{\Omega} \uopt(\xx) \exp(-i \kk \cdot \xx) \,\, d\xx = \bvec{0} \quad \text{for } |\kk|^2 > \ksq \, .
\end{equation}
Including the variational derivative of \cref{truncate2} in the Euler-Lagrange equation, \cref{EulerLagrange}, results in simply subtracting a multiple of $\exp(i \kk \cdot \xx)$ from $\uopt$. The correct choice of the Lagrange multiplier is precisely the Fourier coefficient of that mode in $\uopt$. In other words, direct truncation of $\uopt$ precisely solves the optimization problem from Section \ref{sec:opt_fields} with the additional constraint that the velocity field must lie in the space defined by $\kset$.
In the most basic numerical tests with $\ksq=1$, it is unnecessary to truncate $\uopt$ as the dimension of the native function space is manageable. For $\ksq \ge 2$, however, we will project $\uopt$ onto the smaller space defined by $\kset$ in order to render sampling and optimization feasible.

\subsubsection{Dimensionality of the Fourier spaces}
\label{sec:dimensionality}

In our numerical tests, it is important to consider the dimension of the Fourier spaces from which the magnetic and velocity fields are taken.
For example, in the simplest case of $\ksq=1$, the number of permissible wavevectors in $\Bvec$ is $\nmodes=3$, but, since $\Bvec$ is vector-valued, it has substantially more degrees of freedom. As will be shown below, the number of real degrees of freedom in constructing $\Bvec$ is 9, even after accounting for redundancies. Due to the quadratic term, $\jvec \times \Bvec$, the velocity fields $\uopt$ and $\vrand$ reside in an even larger space. As shown below, the dimension of this space is 24, which is already is no trivial task to numerically sample, and the matter gets substantially worse as $\ksq$ increases.

More generally, \cref{tab:modes} shows how the dimensions of these spaces grow as $\ksq$ increases. The third column shows the number of wavevectors, $\nmodes$, used to construct $\Bvec$ for each value of $\ksq$.
For each $\kk \ne \bvec{0}$, the corresponding Fourier coefficient $\Bhat_\kk$ is constructed through \cref{df_mode} using {\em two} independent complex numbers. Here, the solenoidal constraint reduces the required number of complex coefficients in each $\Bhat_\kk$ from three to two, or equivalently, from 6 real degrees of freedom (d.o.f.)~to 4 real d.o.f. Meanwhile, the Fourier coefficient of the constant mode $\Bhat_\bvec{0}$ is selected with three independent real numbers. Translations can be exploited to further reduce the required d.o.f. In particular, translations in $x_1$ and $x_2$ allow one to take $z_1$ in \cref{df_mode} to be real for $\kk = (1,0)$ and $(0,1)$ without losing any generality. Thus, for $\nmodes$ permissible modes, we have $4(\nmodes)-1$ real d.o.f.~for each $\kk \ne \bvec{0}$ and 3 additional real d.o.f.~for $\kk = \bvec{0}$. Substracting the 2 real d.o.f.~from translations yields a total of $4 (\nmodes-1)+3-2 = 4\nmodes -3$ real d.o.f.~in constructing $\Bvec$, as given in column 4 of \cref{tab:modes}.

Counting the degrees of freedom in the untruncated $\uopt$ is less straightforward due to the quadratic term $\jvec \times \Bvec$. Nonetheless, the fifth column of \cref{tab:modes} lists the real dimension of this space tabulated empirically by computing the cross product and counting the non-zero modes. Notice that the real d.o.f. in the untruncated $\uopt$ grows rapidly with $\ksq$.
If $\uopt$ is truncated according to \cref{truncate1}, the counting is straightforward. The number of permissible wavevectors is the same $\nmodes$, but the fact that $\uopt$ is mean free removes the three real d.o.f.~from the constant mode. At the same time, we {\em cannot} exploit translations to remove any degrees of freedom as we did for $\Bvec$, since that would involve an unjustified assumption about the relative phase of $\Bvec$ and $\uopt$. The result is that real d.o.f.~in the truncated $\uopt$ is always one less than that of $\Bvec$, as shown in the sixth column of \cref{tab:modes}. To give an illustrative example, for $\ksq=4$, the untruncated $\uopt$ resides in an 80-dimensional Fourier space, which would be impractical to sample numerically. Truncating the velocity field, however, reduces the dimension to 24, which is feasible.

\begin{table}[]
\centering
\begin{tabular}{c|c|c|c|c|c}
    $\ksq$ &Fourier wavevectors &number of permissible  
    &real d.o.f. in $\Bvec$, &real d.o.f. in $\uopt$, & real d.o.f. in $\uopt$ \\
    &$\kk = (k_1,k_2)$ & wavevectors, $\nmodes$ 
    &$4\nmodes-3$ &no truncation &with truncation \\\hline
    1 & (0,0), (1,0), (0,1)             &3  &9  &24 &8 \\
    2 & (1,1), (1,-1)                   &5  &17 &48 &16 \\
    4 & (2,0), (0,2)                    &7  &25 &80 &24 \\
    5 & (2,1), (2,-1), (1,2), (1,-2)    &11 &41 &136 &40 \\
\end{tabular}
\caption{The maximum square wavenumber $\ksq$ determines the dimension of the spaces in which $\Bvec$, $\uopt$, and $\vrand$ reside. Column 2 lists the additional wavevectors included in the set $\kset$, and column 3 shows a running count $\nmodes$ of these wavevectors. Once redundancies are removed, the number of real degrees of freedom (d.o.f.) in constructing $\Bvec$ is $4\nmodes-3$, as listed in column 4. Due to the quadratic term, $\jvec \times \Bvec$, $\uopt$ possesses significantly more d.o.f., as shown in column 5. However, $\uopt$ can be projected onto a lower-dimensional space through direct truncation. The real d.o.f.~for this truncated field is shown in column 6. In our numerical tests, the random field $\vrand$ is always sampled from the same space as $\uopt$, i.e.~consistent with the choice of whether or not to truncate $\uopt$.
}
\label{tab:modes}
\end{table}


\subsection{First numerical test: verification that $\uopt$ is optimal}
\label{sec:verify}

With the algorithms established, we now conduct our first numerical test. For a prescribed magnetic field, we aim to verify that the velocity field determined by \cref{PDE2} indeed produces the largest possible instantaneous growth rate $\Mdot$ of magnetic energy. We first randomly select a solenoidal, unit-energy field $\Bvec$ using the approach described in Section \ref{sec:sampling}. That is, the complex numbers $z_1$ and $z_2$ in \cref{df_mode} are randomly sampled in an i.i.d.~fashion to obtain a set of divergence-free modes that form $\Bvec$. This field is then normalized to satisfy $M=1$. Next, we use \cref{prov_soln,uopt_normalized} to numerically solve \cref{PDE2} for the optimal velocity field $\uopt$. With $\uopt$ known, we use \cref{Mdot2} to compute the theoretically optimal growth rate $\MdotOpt$. In all computations we set $\Rm^{-1}=0$ (no magnetic diffusivity) and $w=1/2$ (equal weights of energy and enstrophy).

We aim to compare $\MdotOpt$ to values produced by the same $\Bvec$ but randomly selected velocity fields. We select each random divergence-free velocity field $\vrand$ with the procedure described in Section \ref{sec:sampling} and normalize it via the weighted norm, \cref{wnorm}, that combines energy and enstrophy. In sampling $\vrand$, we draw the complex numbers $z_1$ and $z_2$ from a heavy-tailed distribution and apply the rejection sampling described in \ref{sec:sampling}. These measures help ensure a representation of velocity fields producing large magnitude $\Mdot$. Even so, the space of velocity fields is sometimes too large to practically search (see \cref{tab:modes}), motivating the truncation procedure described in Section \ref{sec:truncation}. For the simplest case of $\ksq=1$, truncation is unnecessary, and we simply sample the 24-dimensional space in which $\uopt$ natively resides. For $\ksq \ge 2$, we truncate $\uopt$, and we draw the random fields $\vrand$ from the same truncated space to ensure a fair comparison between $\uopt$ and $\vrand$.

\begin{figure}[htb]
\centering
\includegraphics[width=.99 \linewidth]{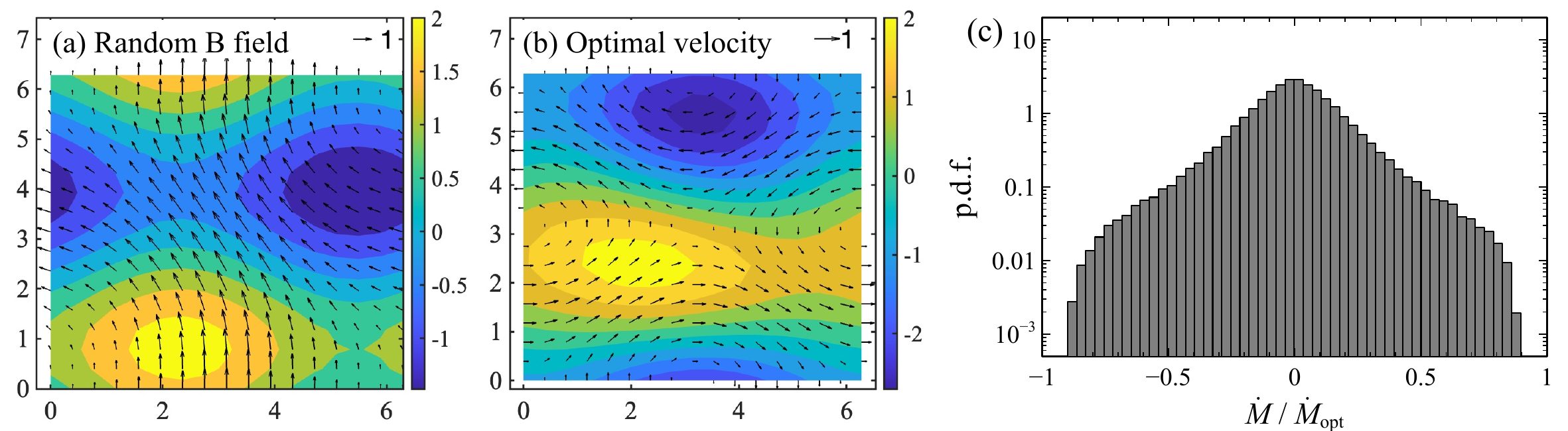}
\caption{Numerical test based on a randomly selected magnetic field of maximal square wavenumber $\ksq=1$.
(a) Visualization of the solenoidal magnetic field $\Bvec$ selected by randomly sampling its Fourier coefficients. The out-of-plane component is shown by color.
(b) Visualization of the corresponding optimal velocity field, $\uopt$, as determined by the numerical solution of \cref{PDE2}. This combination of $\Bvec$ and $\uopt$ yields a magnetic-energy growth rate of $\MdotOpt=1.05$.
(c) Histogram of magnetic-energy growth rates produced by the same $\Bvec$ but randomly selected velocity fields $\vrand$. The histogram shows that $\Mdot < \MdotOpt$ for all sampled velocity fields, confirming that $\uopt$ produces the maximal growth rate.
}
\label{ranB_ksq1}
\end{figure}

For our first test, we consider the simplest case of $\ksq=1$; that is, only the modes corresponding to $(k_1,k_2) = (0,0), (1,0), (0,1)$ and their complex conjugates are permitted in the magnetic field $\Bvec$. We show in \cref{ranB_ksq1}(a) the field $\Bvec$ randomly selected for this test. The figure shows in-plane components $(x_1,x_2)$ of $\Bvec$ with arrows and the out-of-plane component ($x_3$) with color; See the color bar at right. The pictured $\Bvec$ field indeed exhibits random variations, but only large-scale features since the wavenumber is truncated at such a small value.
\Cref{ranB_ksq1}(b) shows the  optimal velocity field $\uopt$, determined by the numerical solution to \cref{PDE2}, that corresponds to this randomly selected $\Bvec$. This field too appears to vary randomly with only large-scale features. The variations in $\uopt$, however, are not truly random as they arise directly from the structure of the given $\Bvec$. This combination of $\Bvec$ and its companion $\uopt$ produces a theoretically optimal growth rate of $\MdotOpt = 1.05$.

\Cref{ranB_ksq1}(c) compares this $\MdotOpt$ against growth rates produced by the same $\Bvec$ but randomly selected velocity fields. To generate the figure, $10^5$ instances of $\vrand$ are sampled with the method described in Section \ref{sec:sampling}, and the normalized values $\Mdot/\MdotOpt$ are compiled into a histogram. 
Observe that the histogram is symmetric about $\Mdot=0$, suggesting that positive and negative growth rates are equally likely if the velocity field is selected randomly. In fact, this symmetry can be seen directly in \cref{Mdot2} (with $\Rm^{-1}=0$) through the mapping $\uu \to -\uu$; that is, for every velocity field producing growth of magnetic energy, there is a corresponding velocity field that produces decay. 
Importantly, \cref{ranB_ksq1}(c) shows  that all sampled values of $\Mdot/\MdotOpt$ are bounded by $\pm 1$, thus confirming that the solution to \cref{PDE2} indeed produces the maximal possible growth rate of magnetic energy.
We remark that setting $\Rm^{-1} > 0$ would yield the same optimizing velocity field $\uopt$, but  lower values of $\Mdot$ overall, since the dissipation term in \cref{Mdot2} is sign-definite and independent of velocity. That is, the entire histogram would shift leftwards by amount $\Rm^{-1} \dashint_{\Omega} {\abs{\jvec}}^2 \, dV$.

\begin{figure}[htb]
\centering
\includegraphics[width=.99 \linewidth]{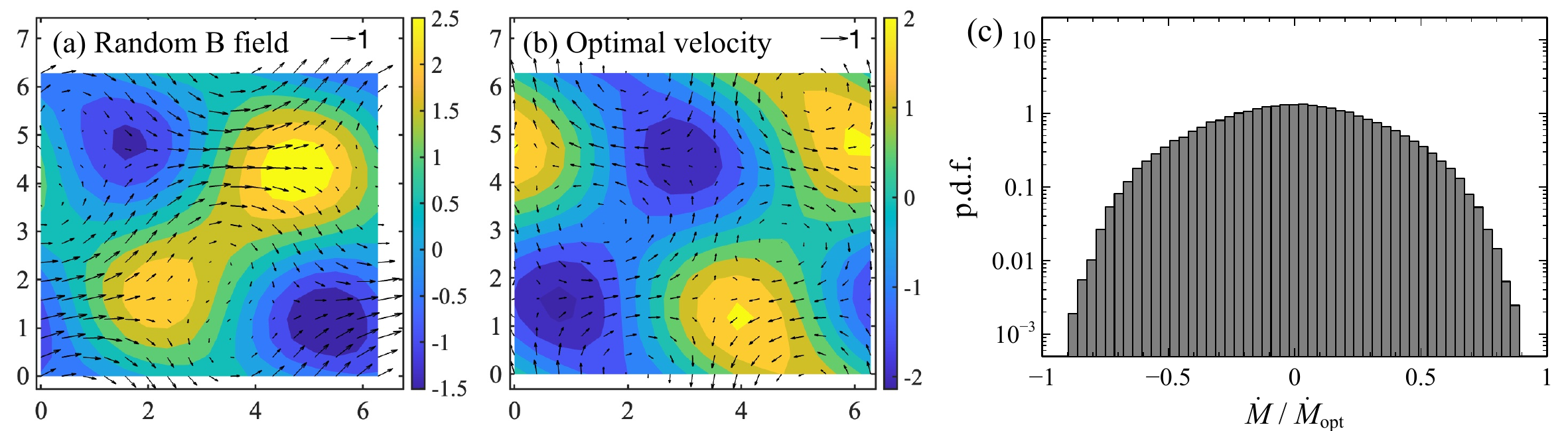}
 \caption{
Numerical test based on a randomly selected magnetic field of maximal square wavenumber $\ksq=2$.
(a) Visualization of the randomly selected $\Bvec$.
(b) Visualization of the corresponding $\uopt$ with spectral truncation applied. This combination of $\Bvec$ and $\uopt$ yields $\MdotOpt=0.61$.
(c) The histogram once again confirms that $\Mdot < \MdotOpt$ for all sampled velocity fields.}
\label{ranB_ksq2}
\end{figure}

In our second numerical test, we set $\ksq=2$ to permit slightly greater spatial complexity. \Cref{ranB_ksq2}(a) shows the magnetic field $\Bvec$ randomly selected for this test, and  \cref{ranB_ksq2}(b) shows the corresponding optimal velocity $\uopt$. In this test, truncation has been applied to $\uopt$ to increase the practicality of searching the velocity-field space. For this particular combination of $\Bvec$ and $\uopt$, the theoretically optimal growth rate is $\MdotOpt = 0.61$.

\Cref{ranB_ksq2}(c) shows the histogram of $\Mdot/\MdotOpt$ values that result from randomly sampled velocity fields. Each field $\vrand$ is sampled from the same space as the truncated $\uopt$. The choice to truncate reduces what would be a 48-dimensional search space to a 16-dimensional one (see \cref{tab:modes}). The main takeaway from this figure is that all sampled values of $\Mdot/\MdotOpt$ lie between $\pm 1$, thus confirming once again that $\uopt$ produces the maximal growth rate possible.

We note that, in the first test, the largest sampled magnitude of $\Mdot$ is roughly 90\% of the optimal value (\cref{ranB_ksq1}(c)), and, in the second test, it is roughly 93\% of the optimum (\cref{ranB_ksq2}(c)). As shown in \cref{tab:modes}, the dimensions of the sample spaces are 24 and 16 respectively (the dimension is smaller in the second test, despite $\ksq$ being larger, because truncation was applied). We note that the additional measures taken in sampling $\vrand$ --- the use of heavy-tailed distributions and rejection sampling --- are crucial to observe samples of $\Mdot$ lying near the optimum. Without these measures, the largest sampled values of $\Mdot$ reach only about 80\% or less of $\MdotOpt$, leaving a significant gap between observations and the theoretical optimum. 

\subsection{Second numerical test: $\uopt$ facilitates numerical optimization of $\Bvec$}
\label{sec:num_optim}

Having verified that \cref{PDE2} produces the optimal velocity field for a given magnetic field, we now turn to the task of numerically optimizing the prescribed magnetic field $\Bvec$ for maximal $\Mdot$. At every stage of the optimization loop, we employ \cref{PDE2} to determine the velocity field $\uopt$ corresponding to the current update of $\Bvec$. This approach enjoys significant advantage over direct numerical optimization of the pair $(\Bvec, \uu)$, since the dimension of the optimization space is roughly half. The tradeoff is one inexpensive numerical solution of \cref{PDE2} per iteration.

In all tests, we perform the optimization over the Fourier coefficients of $\Bvec$ using the multi-start Nelder-Mead algorithm. The dimension of each of these Fourier spaces is given in the fourth column of \cref{tab:modes}. The Nelder-Mead algorithm is a derivative-free simplex method employed in a wide variety of applications \cite{nelder1965simplex, lagarias1998convergence, moore2025search}, and the use of multi-start helps ensure that a global optimum is found \cite{marti2025multi}. Similar to the previous tests, we apply truncation to the velocity fields whenever $\ksq \ge 2$.

\begin{figure}[htb]
\centering
\includegraphics[width=.99 \linewidth]{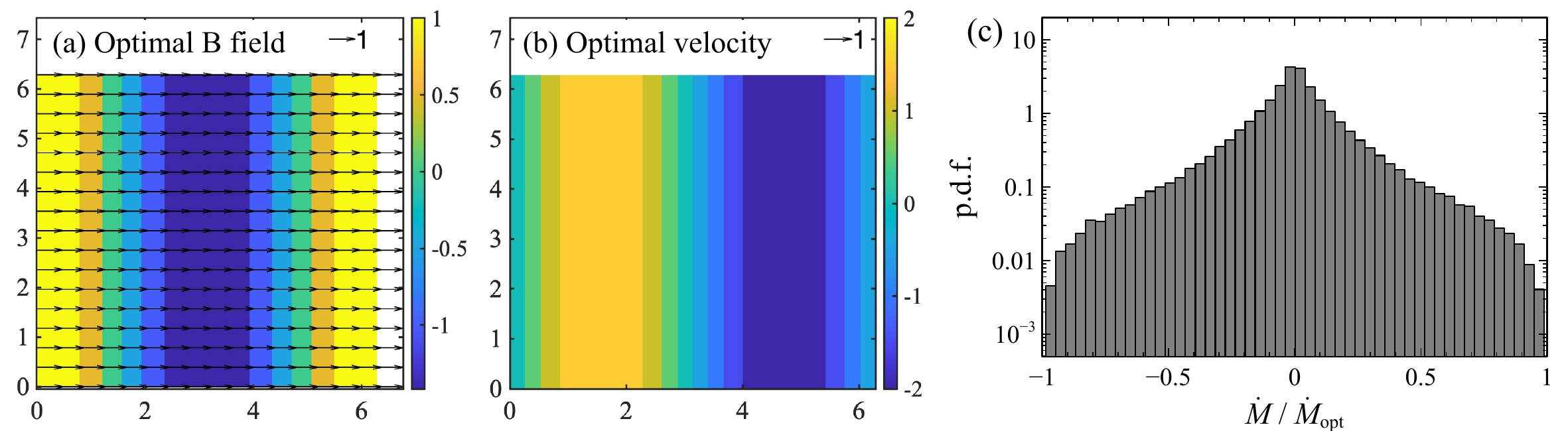}
\caption{
Numerical test with $\ksq=1$ and with $\Bvec = \Bopt$ numerically optimized for maximal $\Mdot$. 
(a) Visualization of the numerically optimized $\Bopt$, and
(b) the companion optimal velocity field $\uopt$. This combination of $\Bopt$ and $\uopt$ produces $\MdotOpt=1.41$.
(c) Histogram of magnetic-energy growth rates produced by the same $\Bopt$ but randomly selected velocity fields. The histogram once again confirms the optimality of $\uopt$. Interestingly, upon optimizing the structure of $\Bvec$, the values of $\Mdot$ produced by randomly selected velocity fields come much closer to the optimal value.}
\label{optB_ksq1}
\end{figure}

For the simplest possible case of $\ksq=1$, \cref{optB_ksq1}(a) shows the magnetic field $\Bopt$ determined by the optimization procedure. This field possesses a remarkably simple structure. It is composed of a constant horizontal component $B_1$ and an oscillatory out-of-plane component $B_3$ that varies only with $x_1$. \Cref{optB_ksq1}(b) shows the corresponding optimal velocity field, $\uopt$, which appears even simpler. In particular, $\uopt$ is a {\em shear flow}, with only one component of velocity that depends on one direction, i.e.~$\uopt = u_3 \ez$ with $u_3 = u_3(x_1)$.  Interestingly, a number of recent studies have highlighted the effectiveness of shear flows in generating dynamo action \cite{blackman2002dynamic, yousef2008generation, heinemann2011large, hughes2013effect, tobias2013shear, squire2015generation, herreman2018minimal, tripathi2026large}, as well as their limitations \cite{proctor2012bounds}. In the case shown in \cref{optB_ksq1}, the velocity shear $u_3(x_1)$ oscillates 90 degrees out of phase with the corresponding magnetic component $B_3(x_1)$.

\Cref{optB_ksq1}(c) shows the histogram of $\Mdot$ values produced by the same $\Bopt$ but randomly selected velocity fields. The histogram once again confirms that, for the $\Bopt$ selected, the companion $\uopt$ produces the highest magnetic-energy growth rate. 
One interesting observation is that, compared to \cref{ranB_ksq1,ranB_ksq2}, the randomly sampled values of $\Mdot$ in \cref{optB_ksq1} come much closer to the bounds $\pm \MdotOpt$.
That is, for a randomly selected $\Bvec$, velocity fields producing near-optimal growth rates are rare. However, for the optimal magnetic field $\Bopt$, near-optimal velocity fields are more abundant in the sampled space. This observation suggests something special about the structure of $\Bopt$ that permits a variety of different velocity fields to strongly amplify, or strongly suppress, its energy.

For this simple case of $\ksq=1$, we have identified simple formulas for the optimizing magnetic and velocity fields. These optimal fields are certainly not unique since the value of $\Mdot$ is invariant to coordinate translations, reflections, and 90-degree rotations. Beyond these simple coordinate transformations, other optimizing fields that attain the same maximum value may exist too. Nonetheless, a representative magnetic field found by our numerical optimization is given by the formula $\Bopt = (1, \, 0, \, \sqrt{2} \, \cos(x) )$. In particular, exactly half of the magnetic energy is contained in the mean field, $\Bhat_{\bvec{0}} = 1 \, \ex$, and the other half in the shear component $\sqrt{2} \cos(x) \, \ez$. The corresponding current is given by $\jvec_{\text{opt}} = (0, \sqrt{2} \sin(x), 0)$. 
We remark that the current is everywhere orthogonal to the magnetic field, $\jopt \cdot \Bopt = 0$, creating conditions for the Lorentz force $\jopt \times \Bopt$ to be optimized. 
The corresponding Lorentz force is given by $\jopt \times \Bopt = (\sin(2x), 0, -\sqrt{2}\sin(x))$ and the corresponding velocity field by $\uopt = (0,0, 2\sin(x))$. The magnetic-energy growth rate that results from the pair ($\Bopt, \uopt$) is $\MdotOpt = \sqrt{2}$. Notice this value is indeed larger than the growth rate $\MdotOpt = 1.05$ from \cref{ranB_ksq1} that was produced by a randomly selected $\Bvec$ (using the same $\ksq$) and its companion optimal velocity.


\begin{figure}[htb]
\centering
\includegraphics[width=.99 \linewidth]{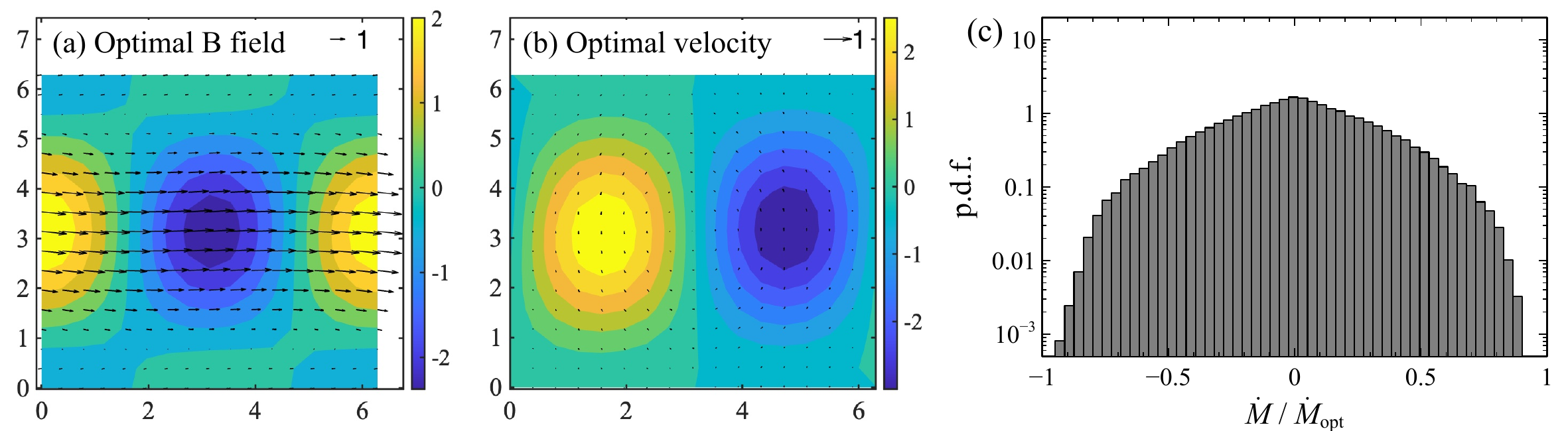}
\caption{
Numerical test with $\ksq=2$ and with $\Bvec = \Bopt$ numerically optimized for maximal $\Mdot$. 
(a) Visualization of the numerically optimized $\Bopt$, and
(b) the companion optimal velocity field $\uopt$. This combination of $\Bopt$ and $\uopt$ produces $\MdotOpt=1.83$.
(c) Histogram of magnetic-energy growth rates produced by the same $\Bopt$ but randomly selected velocity fields. The histogram once again confirms the optimality of $\uopt$.}
\label{optB_ksq2}
\end{figure}

We next perform a similar test with $\ksq=2$. That is, we numerically optimize the magnetic field for maximal $\Mdot$ while using the companion $\uopt$ at every stage of the optimization. \Cref{optB_ksq2}(a)-(b) shows the fields $\Bopt$ and $\uopt$ that result from the procedure. We remark that different seeds in the multi-start routine most often produce fields equivalent to these through translations, reflections, and rotations. Notice that the fields shown in \cref{optB_ksq2} share features with those from \cref{optB_ksq1}. In particular, both tests produce a $\Bopt$ with a strong horizontal mean component and an oscillatory transverse component. Furthermore, the optimal velocity is a shear flow in both tests. The fields shown in \cref{optB_ksq2}, however, exhibit finer spatial features, owing to the fact that higher wavenumbers are permitted. In particular, \cref{optB_ksq2}(a) shows a band running horizontally through the domain that is aligned with the mean-field direction of $\Bvec$. Both the longitudinal, $B_1$, and the transverse, $B_3$, components of the magnetic field are strong within this band and weak outside of it. The longitudinal component is constant along the band, and the energy contained in the mean field, $\Bhat_{\bvec{0}}$, is 1/4 of the total. The transverse component, meanwhile, oscillates along the band, creating the visual appearance of spots. \Cref{optB_ksq2}(b) shows that activity of the velocity field is confined to the same band. Much like the previous case, the shear flow oscillates 90 degrees out of phase with $B_3$.

\Cref{optB_ksq2}(c) shows the histogram of normalized $\Mdot/\MdotOpt$ values produced by the same $\Bopt$ but random velocity fields $\vrand$. The histogram confirms that, compared to the randomly sampled fields, $\uopt$ produces the largest magnetic energy growth rate. The growth rate produced by the combination $(\Bopt,\uopt)$ is $\MdotOpt = 1.83$, which is roughly three times the value $\MdotOpt = 0.61$ from \cref{ranB_ksq2} based on a randomly constructed $\Bvec$ using the same $\ksq$. This comparison illustrates the impact that optimizing the structure of $\Bvec$ has on the potential for magnetic energy growth.

\begin{figure}[htb]
\centering
 \includegraphics[width=.99 \linewidth]{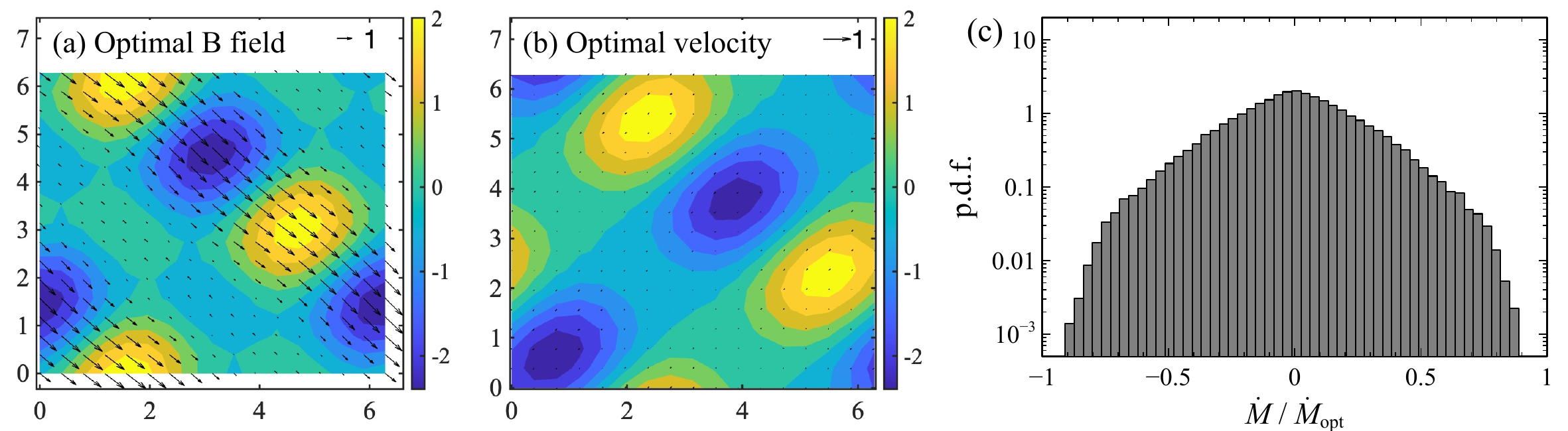}
\caption{
Numerical test with $\ksq=4$ and with $\Bvec = \Bopt$ numerically optimized for maximal $\Mdot$. 
(a) Visualization of the numerically optimized $\Bopt$, and
(b) the companion optimal velocity field $\uopt$. This combination of $\Bopt$ and $\uopt$ produces $\MdotOpt=2.07$.
(c) Histogram of magnetic-energy growth rates produced by the same $\Bopt$ but randomly selected velocity fields. The histogram once again confirms the optimality of $\uopt$.}
\label{optB_ksq4}
\end{figure}

We next perform a similar optimization test with a larger cutoff wavenumber, $\ksq=4$. \Cref{optB_ksq4}(a)--(b) shows the magnetic and velocity fields that result from the optimization, again valid up to translations, reflections, and rotations. Interestingly, the optimal magnetic field bears close resemblance to that from the previous test (\cref{optB_ksq2}), just with finer spatial features. In particular, a narrow band of strong activity crosses the domain and is aligned with the direction of the mean magnetic field. The energy contained in the mean field, $\Bhat_{\bvec{0}}$, is 1/4 of the total, which is precisely the same as in the previous case of $\ksq=2$. The transverse component, $B_3$, oscillates along the band, just as it did in the previous case. \Cref{optB_ksq4}(b) shows that the corresponding velocity field is once again a shear flow, $\uopt = u_3(x_1,x_2) \ez$, that oscillates 90 degrees out of phase with $B_3$. The emergence of a shear flow as the optimizing velocity field in all three tests seems remarkable, especially in this last case of $\ksq=4$ for which there are 24 d.o.f.~in the velocity-field space; i.e.~the possibility of a shear flow arising by chance in such a large space is implausible. 

\Cref{optB_ksq4}(c) shows the histogram of $\Mdot$ values produced by the same $\Bopt$ but randomly selected velocity fields. The histogram confirms, once again, that $\uopt$ produces the highest growth rate. Because this last case of $\ksq=4$ possesses the most d.o.f., it produces the largest maximal growth rate of any of our experiments, $\MdotOpt = 2.07$.

\begin{figure}[htb]
\centering
 \includegraphics[width=.99 \linewidth]{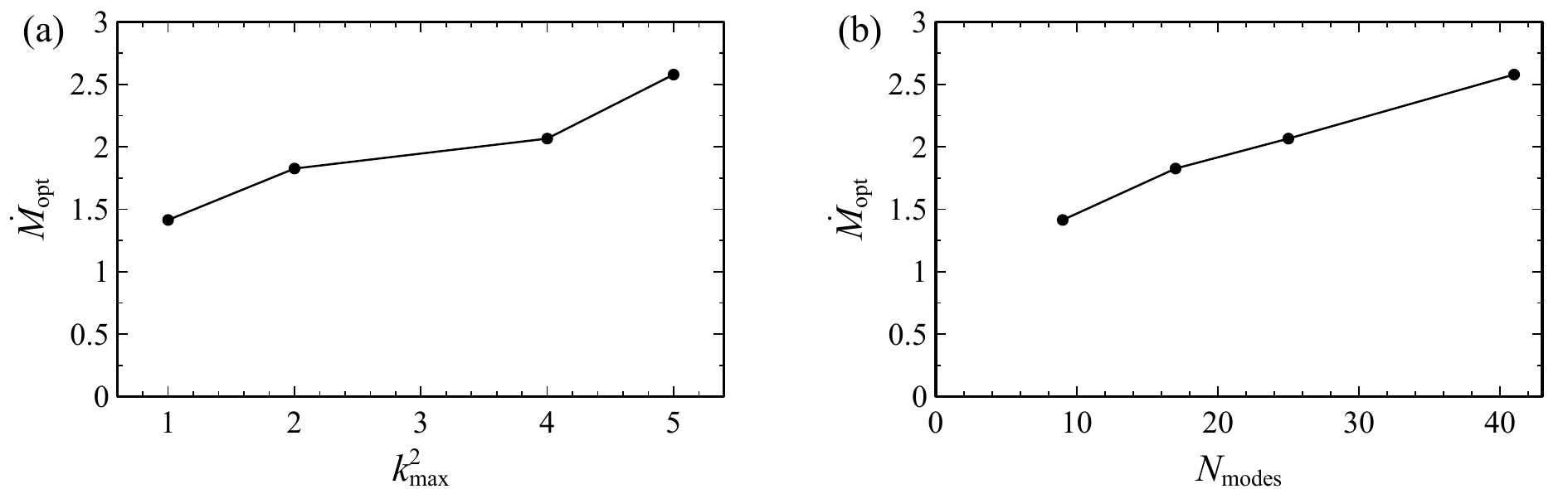}
\caption{
Overall maximum growth rate, $\MdotOpt$, produced by the numerically optimized $\Bopt$ and its companion $\uopt$. (a) Plotting $\MdotOpt$ against $\ksq$ reveals growth, but no simple trend. (b) Plotting $\MdotOpt$ against $\nmodes$ reveals nearly linear growth.
}
\label{fig:Mdot}
\end{figure}


We next examine how the maximum overall growth rate, $\MdotOpt$, grows with the cutoff wavenumber. \Cref{fig:Mdot}(a) shows the values of $\MdotOpt$ --- produced by the numerically optimized $\Bopt$ and its companion $\uopt$ --- plotted against $\ksq$ for values $\ksq=$1, 2, 4, and 5. Although growth of $\MdotOpt$ is evident, there is no obvious trend to the data. However, \cref{fig:Mdot}(b) plots the same $\MdotOpt$ against the number of wavevectors $\nmodes$ and reveals a nearly linear trend. We remark that the last data point, $\ksq=5$, corresponds to $11$ wavevectors and a 41-dimensional space for the numerical optimization of $\Bvec$. Locating a global optimum in a 41-dimensional space with high confidence is a computationally demanding task.
For all smaller cases, $\ksq \le 4$, we have verified convergence of $\MdotOpt$ with respect to the number of specified multi-starts, and our tests suggest 100 multi-starts is sufficient to locate a global maximum. For $\ksq = 5$, we also observe convergence, although it is slower, and roughly 400 multi-starts seem to be required to locate a global maximum. For higher values of $\ksq$, the increased computational burden makes it difficult to draw any firm conclusions regarding the linear trend seen in \cref{fig:Mdot}(b), although it is an interesting question for future research.


\section{Discussion}
\label{sec:discussion}

Here, we have considered a variant of the kinematic dynamo problem in which we treat the seed-magnetic field, rather than the velocity field, as prescribed. Variational calculus leads to a forced Helmholtz PDE, \cref{PDE2}, for the velocity field $\uopt$ that maximizes the instantaneous growth rate of the prescribed magnetic field. This velocity field is jointly constrained in kinetic energy and enstrophy. 
Our numerical experiments of 2.5-dimensional vector fields in doubly-periodic domains confirm that, for any selected magnetic field, $\pm \uopt$ always produces the optimal instantaneous growth rate, $\pm \MdotOpt$, of magnetic energy.
In the near future, we plan to extend these numerical examples to fully 3-dimensional fields in more realistic geometries and with more complex boundary conditions.
We also plan to insert the optimal velocity into dynamic simulations to examine the limits of magnetic-field growth when the velocity field is always selected optimally.

Our numerical experiments illustrate one significant benefit of the newly derived \cref{PDE2}: This equation can substantially accelerate numerical optimization of the magnetic-velocity field pair, $(\Bvec, \uu)$, for maximal $\Mdot$ by reducing the dimension of the optimization space by roughly half. The price to pay for this reduction is small, namely one numerical solution of the linear PDE, \cref{PDE2}, per iteration. 
Alternatively, the eigenvalue approach of the classical kinematic dynamo problem could be employed in optimization, and it would also reduce the search-space dimension by roughly half. The approach outlined here, though, is somewhat more computationally efficient. Given a total of $N$ grid points, computing the appropriate eigenfunction by power iteration would require $O(K N^2)$ operations, where $K$ is the number of iterations required to achieve a specified error tolerance. However, numerically solving the linear PDE \cref{PDE2} with the direct Fourier method outlined here requires $O(N^2)$ operation, with a modest prefactor and no iteration dependence. For larger $N$, a pseudo-spectral method could be used to further reduce the cost to $O(N \log N)$. One interesting possibility for future study is to {\em combine} the eigenvalue approach of the kinematic dynamo problem with the PDE-based approached outlined here to further accelerate optimization. That is, for a given $\Bvec$, the numerical solution to \cref{PDE2} furnishes the optimal velocity field, and for a given $\uu$, the numerical solution to the eigenvalue problem furnishes the optimal magnetic field. Alternating these two methods may prove an effective way to optimize the pair $(\Bvec,\uu)$ without the need for an external optimization algorithm (e.g. a Nelder-mead or quasi-Newton method).

The numerical experiments conducted here reveal the main characteristics of doubly-periodic, 2.5-dimensional fields, $(\Bopt, \uopt)$, that optimize instantaneous magnetic-energy growth. In particular, activity of both the magnetic and flow field is confined to a band that crosses the domain and aligns with the mean magnetic-field component. In the cases $\ksq=2$ and 4, we observe that exactly 1/4 of the total magnetic energy is contained in the mean field. This observation may be meaningful, but we currently do not have a theoretical explanation for it. The optimal magnetic field, in addition to its longitudinal component, exhibits a transverse component $B_3$ that oscillates along the band, thus creating the appearance of spots. The corresponding velocity field is a {\em shear} flow, with transverse component $u_3$ that oscillates 90 degrees out of phase with $B_3$. Outside of the band, both the magnetic and velocity fields are dormant.
It is somewhat surprising that shear flows emerge in all of our optimization tests, even in cases where the optimization space is enormous and shear flows constitute an exceedingly small portion of it. This observation suggests a special role played by shear flows, at least in the case of planar 2.5-dimensional flows, and may relate to recent studies that examine dynamo action created by large-scale mean shear flows \cite{blackman2002dynamic, yousef2008generation, heinemann2011large, hughes2013effect, tobias2013shear, squire2015generation, herreman2018minimal, tripathi2026large}, even though limitations of long-time dynamo action by shear flows are known to exist \cite{proctor2012bounds}.
In the near future, we plan to extend our experiments to examine optimal fields that emerge in fully-three dimensional and spherical settings.

Lastly, the problem studied here may constitute a first step towards inferring fluid flow structures from surface magnetic-field measurements. That is, in some cases, the measurement of surface magnetic fields may be more practical than the direct measurement of flows deep within the interiors of planets or stars \cite{donati2009magnetic, donati2006surprising, gizon2020meridional, stejko2022constraining}. One example is the solar dynamo, where the surface magnetic field has been measured to higher accuracy than has the meridional circulation \cite{charbonneau2014solar, chen2017comprehensive, gizon2020meridional, fuentes2024assessing}.
While the detailed structures of these interior flows are not fully known, the flows must be capable of transferring considerable energy to the magnetic field in order to support it against ohmic dissipation. Thus, a natural starting point in searching for such flows is to determine the velocity field that maximizes energy transfer to the magnetic field, as has been examined here. In future work, we hope to extend these results to dynamic velocity fields, such as those inspired by thermally-convective flows that can reverse their primary direction of circulation chaotically or periodically \cite{moore2024large, zhang2025low}.

\section{Acknowledgment}
N.J.M.~recognizes support from the Office of Naval Research, Grant Number N000142412617, Grant Monitor Dr.~Reza Malek-Madani. We would like to thank Steve Tobias, Keaton Burns, Loren Matilsky, Florentin Daniel, Adrian Fraser, and Laurette Tuckerman for insightful conversations. We would like to also acknowledge the Geophysical Fluid Dynamics program at the Woods Hole Oceanographic Institution for making this collaboration possible.

\bibliographystyle{plain}
\bibliography{bib_dynamo}
\end{document}